\documentclass[conference]{IEEEtran}
\IEEEoverridecommandlockouts
\usepackage{cite}
\usepackage{amsmath,amssymb,amsfonts}
\usepackage{algorithmic}
\usepackage{graphicx}
\usepackage{textcomp}
\usepackage{xcolor}
\usepackage{float}
\usepackage{url}
\def\BibTeX{{\rm B\kern-.05em{\sc i\kern-.025em b}\kern-.08em
    T\kern-.1667em\lower.7ex\hbox{E}\kern-.125emX}}
    
\newcommand{\q}{\mathsf{q}}
\renewcommand{\d}{\mathsf{d}}
\newcommand{\dq}{\mathsf{dq}}

\newcommand{\DQ}{\mathsf{DQ}}

\usepackage{tikz}
\usepackage{pgfplots}
\usepackage{multirow}
\pgfplotsset{compat=newest}
\usetikzlibrary{plotmarks}
\usetikzlibrary{arrows.meta}
\usetikzlibrary{decorations.markings}
\usepgfplotslibrary{patchplots}

\tikzset
{every pin/.style = {pin edge = {<-}},    
> = stealth,                            
flow/.style =    
{decoration = {markings, mark=at position #1 with {\arrow{>}}},
postaction = {decorate}
},
flow/.default = 0.5,          
main/.style = {line width=1pt}                    
}

\begin{document}
\title{Small-Signal Stability Impacts of Load and Network Dynamics on Grid-Forming Inverters
\thanks{This research was supported  by the U.S. Department of Energy by the Advanced Grid Modeling program within the Office of Electricity, under contract DE-AC02-05CH1123, and by the Solar Energy Technologies Office through award 38637 (UNIFI Consortium).}
}

\author{\IEEEauthorblockN{Rodrigo Henriquez-Auba, Jose Daniel Lara}
\IEEEauthorblockA{\textit{Grid Planning and Analysis Center} \\
\textit{National Renewable Energy Laboratory}\\
Golden, CO, USA \\
\{rodrigo.henriquezauba, josedaniel.lara\}@nrel.gov}
\and
\IEEEauthorblockN{Duncan S. Callaway}
\IEEEauthorblockA{\textit{Energy and Resources Group} \\
\textit{University of California Berkeley}\\
Berkeley, CA, USA \\
dcal@berkeley.edu}
}

\maketitle

\begin{abstract}
This paper presents several stability analyses for grid-forming inverters and synchronous generators considering the dynamics of transmission lines and different load models. Load models are usually of secondary importance compared to generation source models, but as the results show, they play a crucial role in stability studies with the introduction of inverter-based resources. Given inverter control time scales, the implications of considering or neglecting electromagnetic transients of the network are very relevant in the stability assessments. In this paper, we perform eigenvalue analyses for inverter-based resources and synchronous machines connected to a load and explore the effects of multiples models under different network representations. We explore maximum loadability of inverter-based resources and synchronous machines, while analyzing the effects of load and network dynamic models on small-signal stability. The results show that the network representation plays a fundamental role in the stability of the system of different load models. The resulting stability regions are significantly different depending on the source and load model considered.
\end{abstract}


\section{Introduction}

As power systems move towards a larger share of renewable energy sources, more inverter-based resources (IBRs) replace traditional generation interfaced via synchronous machines (SMs), creating new challenges to the understanding of system stability, and dynamic behavior \cite{milano2018foundations}. There have been significant efforts to understand the effects of control strategies of IBRs on system stability, e.g., grid-following (GFL) and grid-forming (GFM) using quasi-static Phasors (QSP) and electromagnetic transients (EMT) approaches.
Recent studies have uncovered interactions between excitation systems and inverter controllers using small-signal analyses \cite{markovic2021understanding}, as well as effects of QSP network modeling in the global and small-signal stability of GFM inverters \cite{gross2019effect, henriquez2020grid}.

However, most of IBR system-wide analyses commonly use constant impedance load models despite the fact that load representation has an important influence on dynamic behavior and system stability \cite{concordia1982load}. 
For example,  \cite{kosterev1999model} identifies that accurate load modeling is an important issue in replicating measurements after disturbances. As discussed by Charles Concordia in the 80's and corroborated in a survey in 2010 by the CIGRE working group C.4605 \cite{milanovic2012international}, the industry acknowledges the importance of adequate load modeling. Yet, the emphasis is mostly placed on the accurate modeling of generating units, such as detailed control structures of IBRs, while load models are regarded as of secondary importance, due to lack of information. Further, as power electronics also become commonplace on the load side, the commonly used assumptions about load model structures also need to be revised \cite{arif2017load, emadi2006constant}.  

The main objective of this work is to explore the interaction of static and dynamic load representations and network dynamics on the small-signal stability of GFM inverters. In this paper, we present a detailed study of the effects of modeling assumptions on grid-forming stability and dynamic behavior of GFM IBRs. We focus our analysis in three GFM control strategies: droop, virtual synchronous machine (VSM) and dispatchable virtual oscillator control (dVOC). The results showcase the control interactions depending on the network and load representations in the local stability characteristics. 

The main contributions of this work are:
\begin{itemize}
    \item A stability analysis for network dynamics represented in QSP/EMT domain against constant power and ZIP loads. 
    \item A thorough small-signal stability and bifurcation analysis for maximum loadability scenarios for GFM controls considering different load models. 
    \item A discussion of the increased modeling details required to capture constant power load effects and composite loads in the EMT domain.
\end{itemize}

\textbf{Notation:} We denote the synchronous rotating reference frame (SRF) of the network as $\DQ$, rotating at a fixed frequency $\omega_s = 1$ p.u. 
Bold lowercase symbols are used to represent complex variables in the $\dq$ or $\DQ$  reference frames, $\boldsymbol{x} = x_\d + jx_\q$. Bold capital symbols are used to denote complex matrices $\boldsymbol{Y}$. We use $\dot{x} = dx/dt$ and $f_x = \partial f / \partial x$.  

\begin{figure}[t]
        \centering
        \begin{tikzpicture}[scale = 0.9]
        \pgfplotsset{ticks=none}
        \begin{axis}[%
        width=5cm,
        height=4cm,
        scale only axis,
        xmin=0,
        xmax=2,
        xlabel style={font=\color{white!15!black}},
        ymin=0,
        ymax=2,
        ylabel style={font=\color{white!15!black}},
        ylabel={$v$},
        every axis y label/.style={at={(current axis.north west)},left=0mm},
        axis x line=middle,
        axis y line=left,
        axis line style={-latex}]
       \addplot[red,domain=0.02:1.99, samples=200]{0.3/x};
       \addplot[blue,domain=0.0:1.99, samples=200]{1 - (x/5)^2};
       \addplot +[black, mark=none, dashed] coordinates {(0.3, 0) (0.3, 1)};
       \addplot +[black, mark=none, dashed] coordinates {(0.42, 0) (0.42, 1)};
       \addplot +[black, mark=none, dashed] coordinates {(0.18, 0) (0.18, 1.7)};
       \draw [<->](0.3, 0.2) -- (0.42, 0.2) ;
       \draw [<->](0.3, 0.2) -- (0.18, 0.2) ;
        \end{axis}
        \node[below] at (4.8,0) {$i$};
        \node[above] at (1.2, 2) {\footnotesize$(i_0,v_0)$};
        \node[above] at (1.25, 0.25) {\scriptsize$\Delta i$};
        \node[above] at (4.25, 1.78) {\scriptsize Source};
        \node[above] at (4.25, 0.35) {\scriptsize CPL};
        \node[] at (0.75,1.99) {$\bullet$};
        \node[blue] at (1.05, 1.97) {$\bullet$};
        \node[red] at (1.05, 1.41) {$\bullet$};
        \end{tikzpicture}
        \caption{$v$-$i$ curves for common voltage sources and ideal CPLs.}
        \label{fig:vi_curves}
    \end{figure}
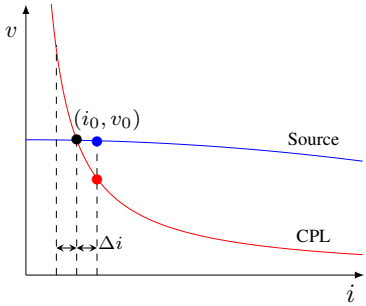

\section{Instability aspects for Constant Power Loads \label{sec:cpl_inst}}

Constant power loads (CPLs) are one of the most ubiquitous load models used in power flow studies. As a starting point, we give intuition about the stability challenges in supplying CPLs, first for DC systems, on which a voltage source is supplying a CPL through an inductor. This intuition is similar for dynamic phasor analysis in AC systems, as we discuss later. 
As mentioned in \cite{emadi2006constant}, \textit{ideal} CPLs satisfy instantaneously at all times $v\cdot i = \text{constant}$; thus, if a voltage across a CPL increases/decreases, then the current must decrease/increase respectively to satisfy the constant power condition. This behavior can be visualized in Figure \ref{fig:vi_curves}. 
Most generation sources, such as GFM IBRs or generators whose Automatic Voltage Regulators (AVRs) are set to voltage control, describe a concave $v$-$i$ curve that decreases the voltage when the current increases. The intersection $x_0 = (i_0,v_0)$ (black circle) determines the operating point of the source connected to the load. Suppose that an external perturbation increases the current $\Delta i$, then the load voltage (in red) is lower than the source voltage (in blue), resulting in an increase of the current provided by the source, while a similar behavior occurs if a disturbance decreases the current. In both cases, the system moves away from $x_0$, resulting in an unstable operating condition.

Network dynamics also play a role in the stability assessment of AC systems with CPLs. When network dynamics are neglected, as in standard QSP modeling, the circuit dynamics are represented by the algebraic relationship using the admittance matrix $\boldsymbol{i} = \boldsymbol{Y}\boldsymbol{v}$. In this case, CPLs do not necessarily induce instability in the dynamic model of a generation system. However, using EMT dynamic phasor modeling, Allen \& Ili\'c \cite{allen2000interaction} show that small-signal stability for a source connected to a CPL ($S = \boldsymbol{v}\boldsymbol{i}^* =$ constant) via a series resistor and inductance requires the condition: $\left|\boldsymbol{v}_\text{source} - \boldsymbol{v}_\text{load}\right| > \left|\boldsymbol{v}_\text{load}\right|$. This condition is highly unlikely to occur in AC transmission systems, where $|\boldsymbol{v}_\text{source}| \approx |\boldsymbol{v}_\text{load}| \approx 1.0$ p.u. Thus, the operating points over the desired part of the P-V curve on typical load conditions are always unstable. In practice this is not observed, since load models and composition changes the stability regions, as we discuss next.


\section{Maximum loadability analysis \label{sec:loadability}}

\begin{figure*}[t]
    \centering
    \includegraphics[width=0.7\textwidth]{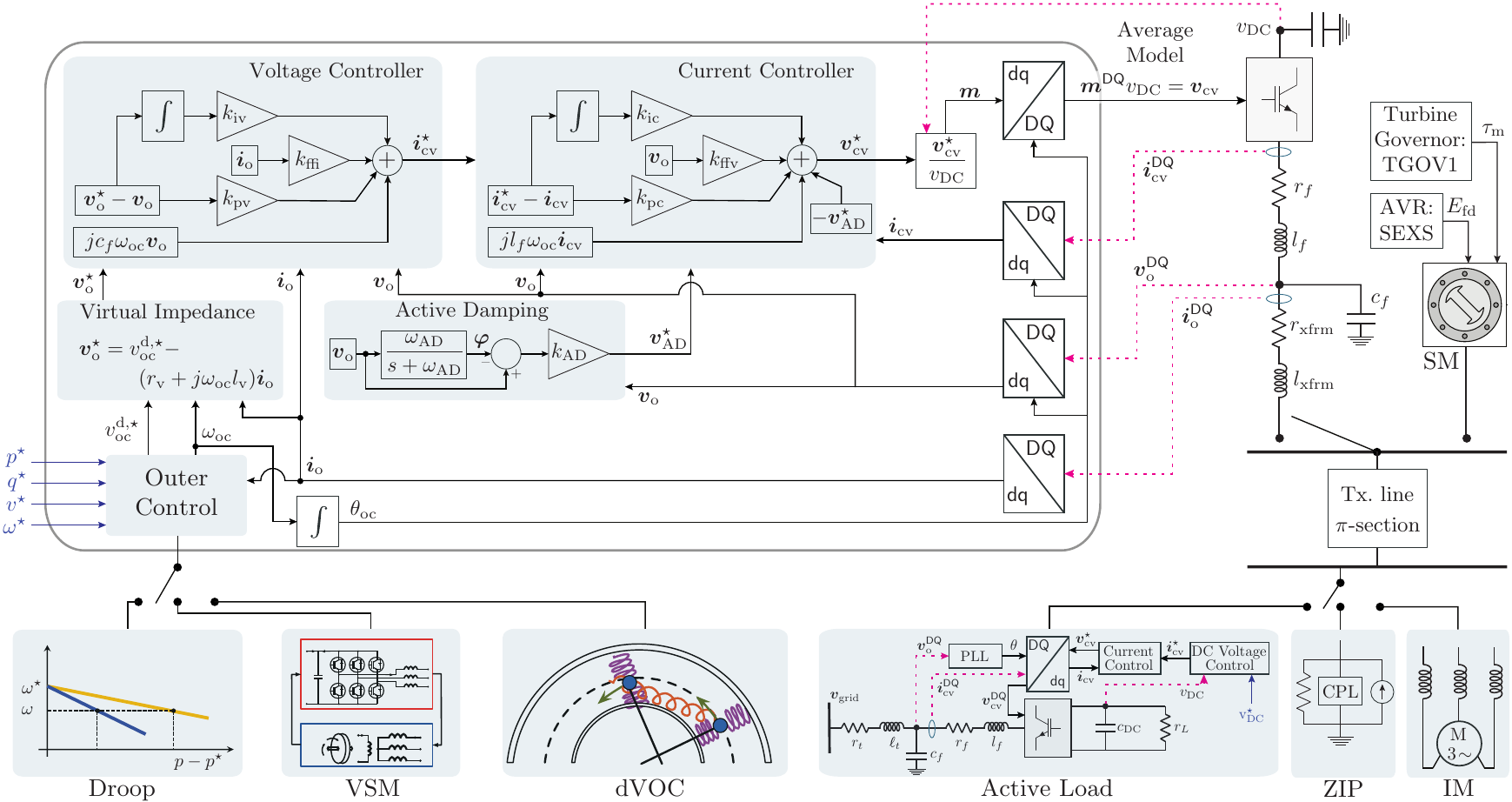}
    \caption{Problem set-up for load studies.}
    \label{fig:setup}
\end{figure*}

To study the effect of load modeling and GFM controls on system stability, we consider a two-bus source vs load balanced system as depicted in Figure \ref{fig:setup}. For GFM inverters, we use a detailed outer control model and inner control models taken from \cite{d2015virtual, ajala2021model, markovic2021understanding}, and tune the outer control parameters to have a 2\% $P$-$\omega$ p.u./p.u. and 5\% $Q$-$v$ p.u./p.u. droop steady-state response using the procedure described in \cite{johnson2022generic}. Since the focus is on small-signal stability, we ignore hybrid dynamical effects of limiters. 

For comparison purposes, the simulation also includes results for SM models. We utilize the GENROU machine model for QSP studies and the Marconato machine model for EMT studies \cite{milano2010power}. To study voltage instabilities it is common to analyze the power/voltage (P-V) curves to understand if the theoretical maximum value of power can be delivered through a single line. From a static load flow perspective, this maximum value is attained at the \textit{nose} of the upper part of the P-V curve. However, as our results will show, network and load dynamics reduce the stability regions. The observed bifurcation is different depending on the dynamic model. When a purely real eigenvalue crosses into the positive plane, it produces a transcritical bifurcation; if a pair of complex conjugate eigenvalue crosses to the positive plane it is known as a Hopf bifurcation, and a singularity-induced bifurcation (SIB) implies that an eigenvalue wraps around $\infty$. From a practical perspective the bifurcation informs which dynamics to monitor for instabilities. For example, in the case of a pair of complex conjugate eigenvalues, it is expected to observe undamped oscillations at the eigenvalues frequency. The reader is referred to \cite{khalilbook, marszalek2005singularity} for more details on bifurcation analysis.

Eigenvalue analyses and time-domain simulations are implemented using the Julia package \texttt{PowerSimulationsDynamics.jl} and  available in the Github repository \url{https://github.com/Energy-MAC/GFM-LoadModeling}. Summary of the main results of the following subsections are presented in Table \ref{tab:results}.

\begin{table}[t]
\centering
\caption{Summary of Maximum Loadability results}
\label{tab:results}
\begin{tabular}{|c|c|c|c|c|}
\hline
Case                                                                           & \begin{tabular}[c]{@{}c@{}}Source\\ Model\end{tabular} & $P^\star$ {[}p.u.{]} & Bifurcation                                                    & \begin{tabular}[c]{@{}c@{}}Participating \\ States\end{tabular} \\ \hline \hline
\multirow{3}{*}{QSP CPL}                                                       & GENROU                                                 & 1.169              & Hopf                                                           & $e_\q$, $E_\text{fd}$                                           \\ \cline{2-5} 
                                                                               & Droop/VSM                                              & 1.242              & \begin{tabular}[c]{@{}c@{}}Singularity-\\ Induced\end{tabular} & $\boldsymbol{\xi}_{\dq}$                                        \\ \cline{2-5} 
                                                                               & dVOC                                                   & 1.241              & \begin{tabular}[c]{@{}c@{}}Singularity-\\ Induced\end{tabular} & $\boldsymbol{\xi}_{\dq}$                                        \\ \hline
EMT CPL                                                                        & All                                                    & 0.0    & Unstable                                                       & \begin{tabular}[c]{@{}c@{}}See\\ Sec. \ref{sec:cpl_inst}\end{tabular}          \\ \hline
\begin{tabular}[c]{@{}c@{}}QSP/EMT\\ CIL or CCL\end{tabular}                   & All                                                    & 4.5                & Stable                                                         &                                                                 \\ \hline
\multirow{3}{*}{\begin{tabular}[c]{@{}c@{}}EMT Single\\ -Cage IM\end{tabular}} & Marconato                                              & 0.688              & Hopf                                                           & $e_\q$, $E_\text{fd}$                                           \\ \cline{2-5} 
                                                                               & Droop/VSM                                              & 1.015              & Transcritical                                                  & $\boldsymbol{\psi}_{\dq}$                                       \\ \cline{2-5} 
                                                                               & dVOC                                                   & 1.049              & Transcritical                                                  & $\boldsymbol{\psi}_{\dq}$                                       \\ \hline
\multirow{3}{*}{\begin{tabular}[c]{@{}c@{}}EMT \\ Active Load\end{tabular}}    & Marconato                                              & 1.075              & Hopf                                                           & $e_\q$, $E_\text{fd}$                                           \\ \cline{2-5} 
                                                                               & Droop/VSM                                              & 1.458             & Hopf                                                           & $\zeta$, $v_\text{DC}$                                          \\ \cline{2-5} 
                                                                               & dVOC                                                   & 1.430             & Hopf                                                           & $\zeta$, $v_\text{DC}$                                          \\ \hline
\end{tabular}
\end{table}

\subsection{QSP P-V curves for CPLs} \label{sec:pv_qsp}

A QSP analysis ignores network dynamics and assumes that circuit dynamics evolve to a stable equilibrium. 
Thus, filters and transmission lines are modeled using the algebraic admittance matrix representation $\boldsymbol{i} = \boldsymbol{Y}\boldsymbol{v}$. 

Figure \ref{fig:pv_curve_together}-(a) depicts our load flow (from 0 to 4.5 pu, using a step of 0.01 pu) and eigenvalue analysis in the QSP domain assuming a single source connected to a CPL. Additional P-V curves for all models are available in the repository. 

We examined the eigenvalues of the linearized reduced system $A = f_x - f_y g_y^{-1} g_x$ (see Section \ref{sec:composite}) at each operating point. For all source models, the colored regions depict stable operating points, while empty regions showcase unstable operating conditions. We can observe that the instability boundary occurs at similar loading levels for all generation sources $P \approx 1.2$ p.u. (see Table \ref{tab:results}); however, the bifurcation characteristics are significantly different between GFM inverters and the GENROU model. In the GENROU case, the critical eigenvalues are a complex pair that crosses the $j\omega$ axis when the load reaches the critical $P^\star$. This behavior corresponds to a subcritical \textit{Hopf bifurcation} for SMs \cite{chow1995systems}, where the $\q$-axis voltage $e_\q'$ and the field voltage $E_\text{fd}$ states have most significant participation factors in the unstable eigenvalues.

The bifurcation characteristics of the GFM inverters are significantly different than the SM case. Specifically, we observe a SIB as the load value increases \cite{marszalek2005singularity}. When the system reaches $P^\star$ one eigenvalue, associated with the integrator of the GFM voltage controller $\boldsymbol{\xi}_\dq$, crosses \textpm $\infty$ to the other side of the complex plane, changing the stability of the system.

Furthermore, the stability regions of the Droop/VSM are equivalent since the outer control parameters achieve the same steady-state response, and the inner control parameters are the same. dVOC results are different since the droop behavior depends on $e_0$, which changes at different load levels \cite{johnson2022generic}, in this case for dVOC parameters we assume $e_0 = 1$ p.u.

\subsection{EMT P-V curves for ZIP and IM models}

The analysis in this section has the same setup as Subsection \ref{sec:pv_qsp}, but now including detailed circuit models for the AC filters and transmission lines in the $\DQ$ network reference frame, described by equations \eqref{eq:current} (see Section \ref{sec:composite}). \textit{Ideal} CPL models result in a unstable system as the eigenvalues associated with the transmission become positive, as discussed in Section \ref{sec:cpl_inst}. On the contrary, algebraic models for constant current load (CCL) and constant impedance load (CIL) result in stable regions for all cases considered. 

The single-cage induction machine (IM) load uses 5th-order dynamic phasor EMT model \cite{krause2013analysis} including a capacitor in parallel to ensure unitary power factor of the IM device. A similar P-V curve is constructed (available in the repository) with the IM parameters. Table \ref{tab:results} showcases that in this case the system exhibits a similar Hopf bifurcation for the states $E_\text{fd}$ and $e_\q'$ for the SM case. These results are similar to the QSP-CPL case, in which the critical eigenvalues are associated with the excitation system. This reveals that AVR re-tuning or the addition of a power system stabilizer can enhance stability regions. Bifurcation analysis informs that a exciter gain $K$ is an initial approach to modify to improve stability.

In contrast to the QSP-CPL setting, there are no algebraic states when the IM is supplied through a GFM, so no SIB can occur. However, for GFM inverters at the critical load $P^\star$, a single real eigenvalue crosses to the right hand side causing a transcritical bifurcation. The states associated with this eigenvalue, according to its participation factors, are the IM rotor flux linkages $\boldsymbol{\psi}_{\dq}^r$. The analysis reflects that inverter current control re-parametrization can reduce interactions with the stator flux states to address the issue. Future work will focus on sensitivity analysis on current control gains to improve stability. In addition, we seek to explore how changing the IM interface to a speed drive could change stability regions by isolating the stator fluxes.

\subsection{EMT P-V curves for an active load}

As discussed in Section \ref{sec:cpl_inst}, \textit{ideal} CPL models are not suitable for EMT studies, so in this section we employ a 12-state model \emph{Active Load} model  that regulates a DC voltage to supply a resistor $r_L$. This model induces a CPL-\emph{like} behavior as it tries to maintain a fixed DC voltage to supply $P = v_\text{DC}^2 / r_L$ \cite{mahmoudi2016new}. Figure \ref{fig:setup} showcases the \emph{Active Load} model, on which the reference term $i_\q^\star$ from the DC voltage controller is chosen to regulate minimum reactive power consumed from the load.

\begin{figure}[t]
    \centering
    \includegraphics[width=0.38\textwidth]{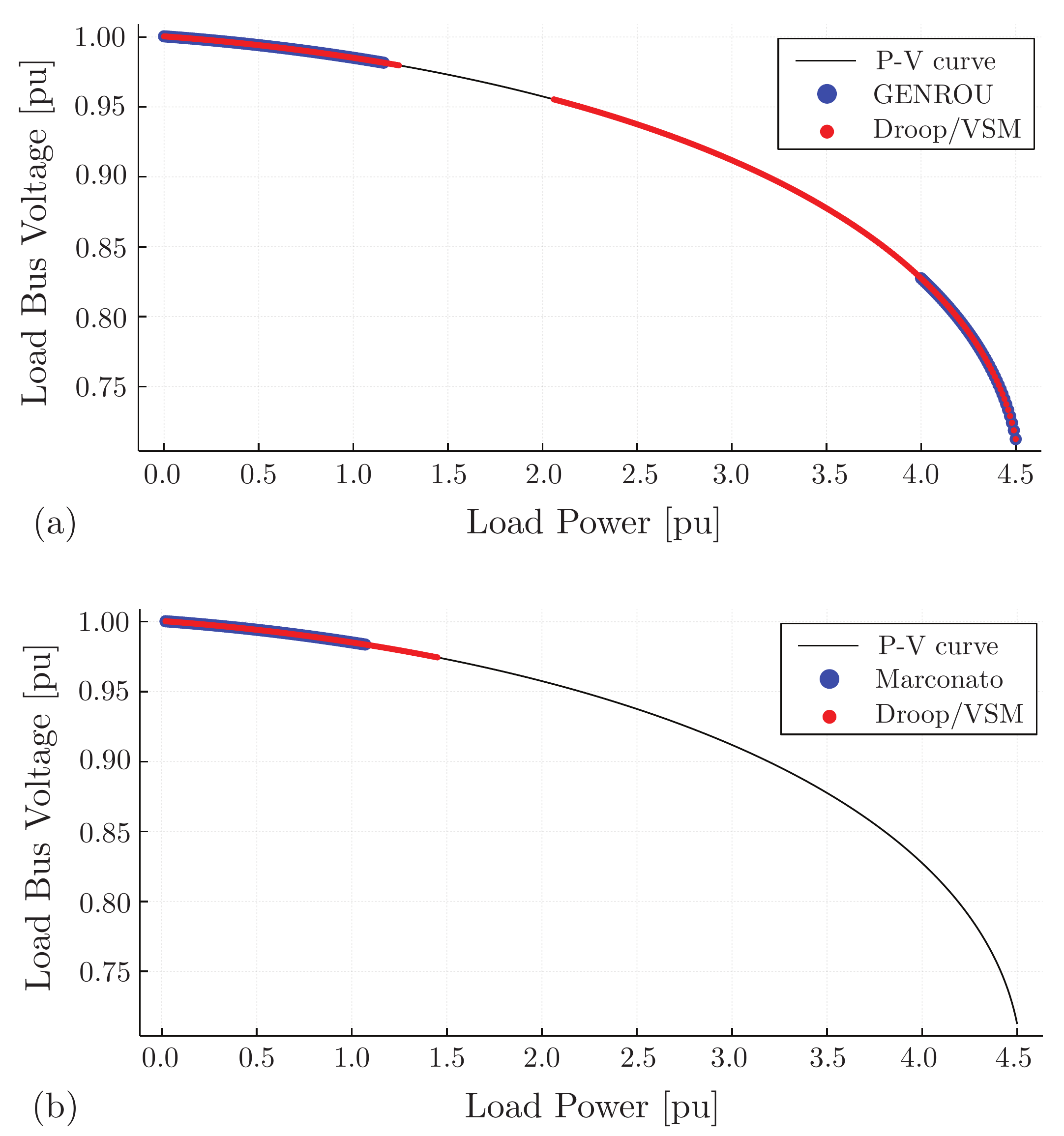}
    \caption{P-V curves for (a) QSP/CPL model, (b) EMT/active load model. Colored regions depict stable operating points, while a black line showcase small-signal unstable operating conditions}
    \label{fig:pv_curve_together}
\end{figure}

\begin{figure}[t]
    \centering
    \includegraphics[width=0.37\textwidth]{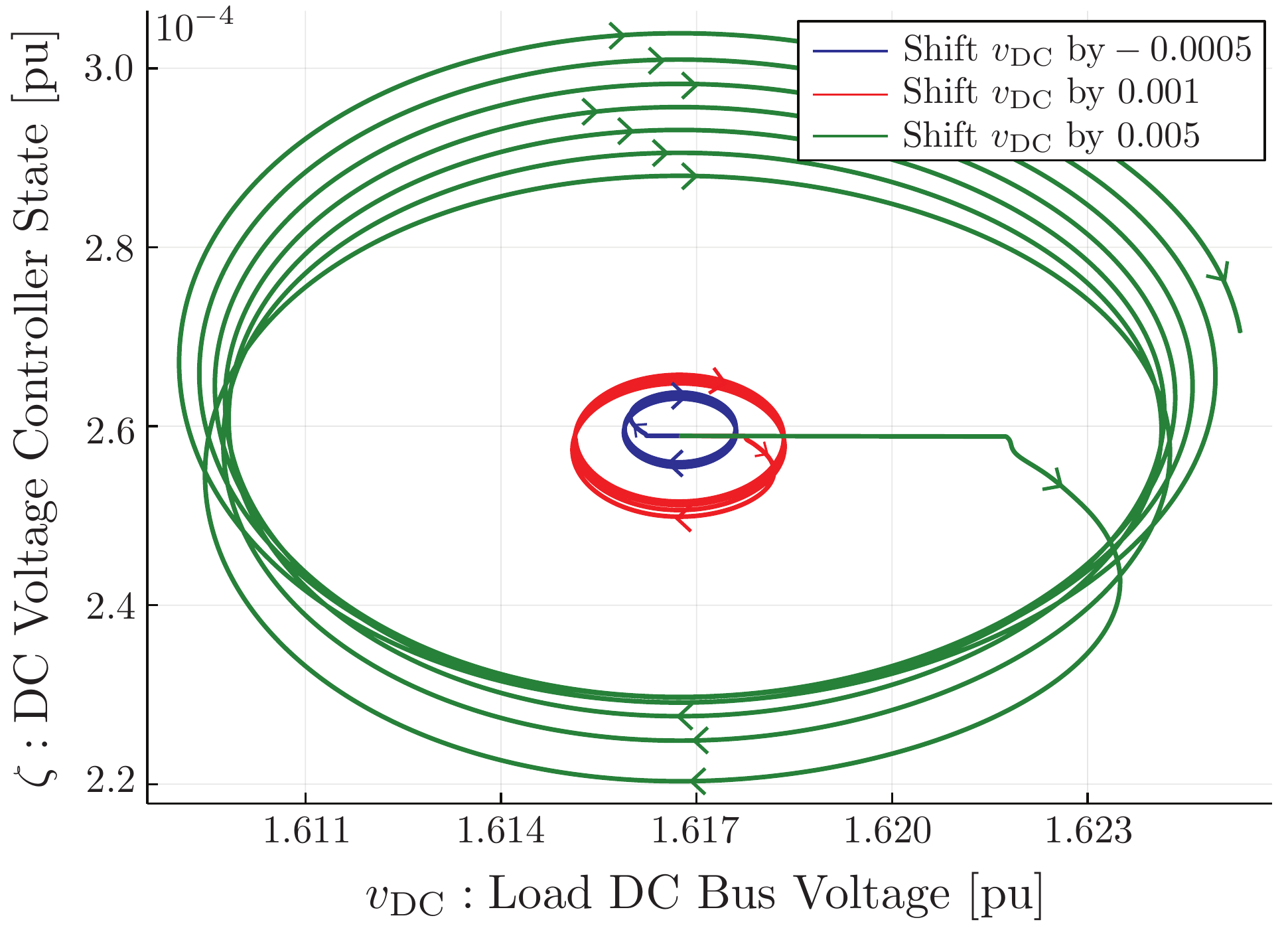}
    \caption{Phase portrait $v_\text{DC}$--$\zeta$ for droop GFM supplying active load at $P^\star$.}
    \label{fig:limit_cycle}
\end{figure}

The resulting P-V curves are depicted in Figure \ref{fig:pv_curve_together}-(b). The GFM inverters case shows two critical eigenvalues associated with the load's model DC voltage controller integrator state $\zeta$ and the DC Voltage $v_\text{DC}$. As the load power is increased, two complex conjugate eigenvalues move from the left-hand to right-hand side complex plane generating a Hopf bifurcation. A phase portrait, depicted in Figure \ref{fig:limit_cycle}, showcases the states $v_\text{DC}$ and $\zeta$, as we introduce different perturbations in the value of $v_\text{DC}$ to assess the system stability. At the critical load $P^\star$ there an is unstable limit cycle around a locally stable equilibrium point, which is consistent with a subcritical Hopf bifurcation. A re-tuning in the active load DC voltage PI controller is a future direction to explore to enhance system stability, however, it is unlikely to be feasible to change load parameters to improve stability, so parameter sweep of inverter parameters will be required as we introduce more complex load models that may not be observed with traditional models, such as the ZIP model described in the next section.

\begin{figure}[t]
    \centering
    \includegraphics[width=0.23\textwidth]{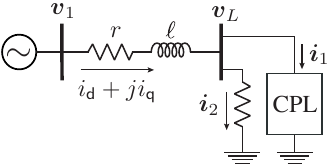}
    \caption{ZIP Load example.}
    \label{fig:cpl_r}
\end{figure}

\section{Importance of Composite Load Modeling in GFM dominated systems \label{sec:composite}}

The ZIP model consists of three loads in parallel, a CPL, a CCL, and a CIL, and it is ubiquitous in QSP power system studies. A ZIP load can capture the static and dynamic behavior of many aggregated composite loads in power systems with the appropriate proportion assignment to each sub-model. Unfortunately, there are no closed-form results (such as those presented in \cite{allen2000interaction}) on small-signal stability for the dynamic phasor domain. However, it is possible to obtain results in some cases; for example, in the system depicted by Figure \ref{fig:cpl_r}, the electric current dynamic phasor (in p.u.) flowing through the line are given by:
\begin{align}
\label{eq:current}
    \frac{\ell}{\Omega_b} \frac{d\boldsymbol{i}_\dq}{dt} &= \boldsymbol{v}_{\dq 1} - \boldsymbol{v}_{\dq L} - r\boldsymbol{i}_\dq - jx \boldsymbol{i}_\dq, 
\end{align}
in which $\Omega_b = 2\pi60$ rad/s the base frequency and $x = \omega_s \ell$ is the line reactance, where $\omega_s = 1.0$ p.u. Given an operating point, we are looking to compute the eigenvalues of the linearized system. To do so we solve for the term $\boldsymbol{v}_L$ given the load parameters of the resistor $r_L$ and CPL values ($P,Q$):
\begin{align}
    \boldsymbol{i} = \boldsymbol{i}_1 + \boldsymbol{i}_2, \qquad \boldsymbol{v}_L\boldsymbol{i}_1^* = P + jQ, \qquad  \boldsymbol{i}_2 = \frac{\boldsymbol{v}_L}{r_L} \label{eq:algs}
\end{align}
  
Equations \eqref{eq:algs} are a system of non-linear equations where is possible to solve $\boldsymbol{v}_L$ as a function of $\boldsymbol{i}$, resulting in two possible solutions. Given the complexity of symbolic solution, there is no closed form relationship for assessing stability as a function of the proportion between CIL $\boldsymbol{i}_1$ and CPL $\boldsymbol{i}_2$ in the ZIP load model. However, it is possible to analyze the stability by constructing the following system of Differential-Algebraic (DAE) equations: $\dot{x} = f(x,y,\eta),~    0 = g(x,y,\eta)$,
where $f(x,y,\eta)$ represents the line current equations \eqref{eq:current} and $g(x,y,\eta)$ are the algebraic equations of the ZIP load \eqref{eq:algs}. In here, $x = \boldsymbol{i}$ are differential states and $y = [\boldsymbol{v}_{ L}, \boldsymbol{i}_{1}, \boldsymbol{i}_{2}]$ are algebraic states and the parameter $\eta$ is used to control the proportion of CPL to CIL. Assuming no reactive power load (i.e. $Q = 0$) in this system, we obtain the determinant of the Jacobian of the algebraic constraints with respect to the algebraic states as follows:
\begin{align}
    \det(g_y(x,y,\eta)) = |\boldsymbol{i}_1|^2 - \frac{|\boldsymbol{v}_L|^2}{r_L^2} = |\boldsymbol{i}_1|^2 - |\boldsymbol{i}_2|^2
    \label{eq:jac_alg}
\end{align}
Equation \eqref{eq:jac_alg} becomes zero when the magnitude of the currents flowing through the CPL and the impedance are equal introducing a \textit{singularity-induced bifurcation} \cite{marszalek2005singularity}. 

We illustrate the bifurcation with the following numerical example: consider the system presented in Figure \ref{fig:cpl_r} with $r = 0.01,~ x = 0.1,~ P_\text{cpl} = 1$ p.u., and $\boldsymbol{v}_1 = 1.0 + 0j$ p.u. The proportion of power flowing through the resistor ($P_{r_L}$) and the CPL ($\eta P_\text{cpl}$) can be adjusted using the parameter $\eta \in (0,1)$ as follows:
\begin{subequations}
\label{eq:load_prop}
\begin{align}
    P_\text{zip} = \eta P_\text{cpl} + P_{r_L} = 1 ~\to~ P_{r_L} &= 1 - \eta P_\text{cpl}\\
    P_{r_L} = \frac{|\boldsymbol{v}_L|^2}{r_L} = 1 - \eta P_\text{cpl} ~\to~ r_L &= \frac{|\boldsymbol{v}_L|^2}{1-\eta P_\text{cpl}}
\end{align}  
\end{subequations}
such that we ensure the total load to be always 1 p.u. (reactive power is set to zero); hence, steady-state voltage is constant at $|\boldsymbol{v}_L| = 0.985$ p.u. 
This allows us to traverse through the singularity manifold $S \equiv \{(x,y,\eta) \mid \det(g_y(x,y,\eta)) = 0\}$, to analyze the trajectory of the line current $i_\d,i_\q$ eigenvalues.  

\begin{figure}[t]
    \centering
    \includegraphics[width=0.48\textwidth]{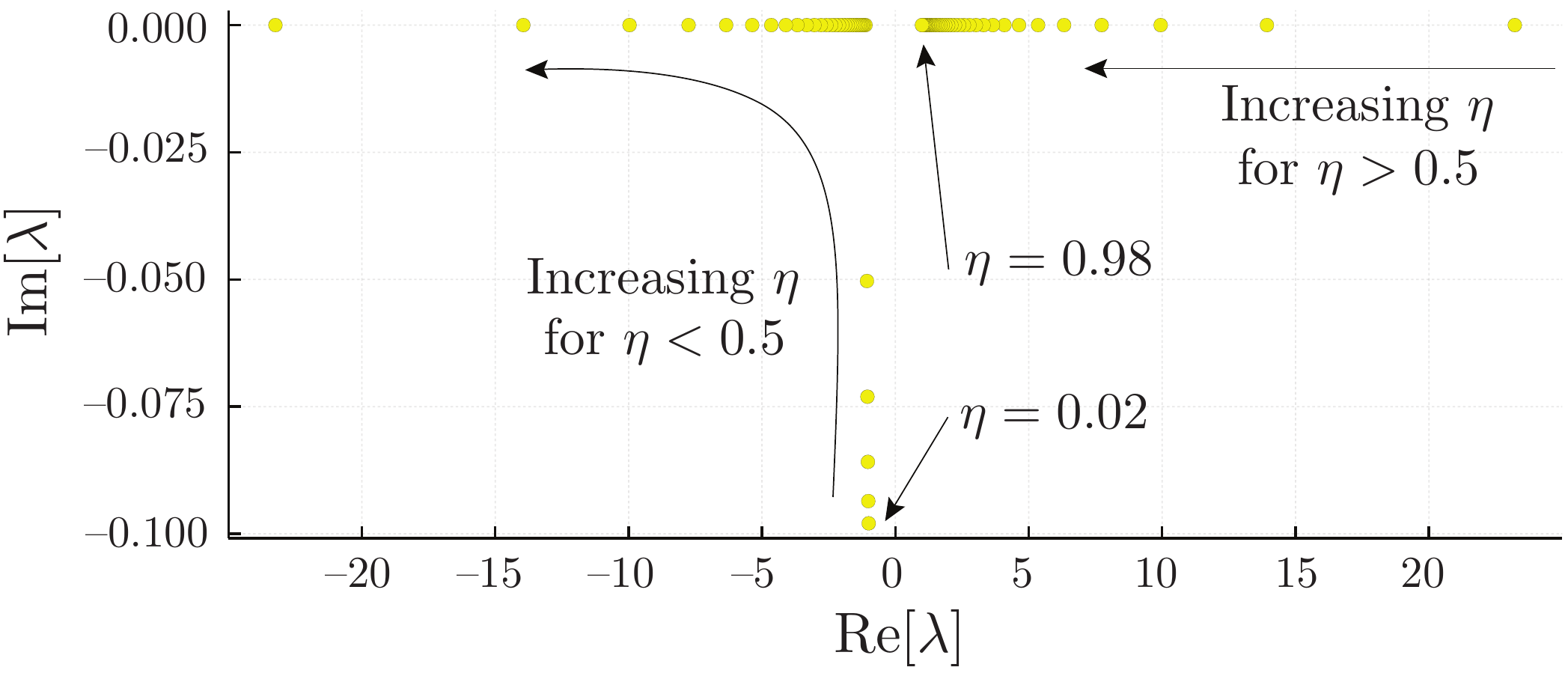}
    \caption{Root locus as $\eta$ changes from $0 \to 1$.}
    \label{fig:eigs_eta}
\end{figure}

As we change the parameter $\eta$ from $0 \to 1$, one of the eigenvalues of the reduced system ($A = f_x - f_y g_y^{-1} g_x$) moves towards --$\infty$, and changes sign as we cross the singularity manifold at $\eta^\star = 0.5$, while the second eigenvalue remains on the left side of the complex plane. The line current states remain small-signal stable if $\eta < 0.5$, a condition where the current magnitude through the resistor is larger than the CPL current magnitude. Figure \ref{fig:eigs_eta} showcases the root-locus of this eigenvalue as we increase $\eta$, on which we observe that diverges through --$\infty$, changing sign when $\eta \ge 0.5$. 

The example highlights the importance of load model composition on network stability. As we move towards a system with larger shares of IBRs and fast dynamics become more relevant, the results show the dependence between current stability and different proportions of a ZIP load model. The analysis showcases the potential inadequacies of modeling only CIL or CPL and the need for detailed and accurate composite load models as we increase the shares of IBRs.

\section{Conclusions}

This paper demonstrates that small-signal stability conclusions for IBRs are heavily dependent on modeling assumptions regarding electromagnetic network phenomena and load representation. The results indicate that common load modeling practices need to be revisited for future dynamic studies in the presence of IBRs. As power systems integrate increasing amounts of electronics-based loads that behave as CPLs and IBRs, more advanced models are needed to assess stability. 

However, we have only explored single-line systems and a specific set of parameters and load models. More comprehensive studies are needed using composite load models in more extensive systems to properly assess the changing stability regions as we increase system loading. Future work should focus on understanding real load dynamics by exploring different load composition scenarios and using a combination of SMs and IBRs controls in the system, while studying parameter tuning on their controls to enhance stability. Further, the inclusion of limiters for SMs controllers and current limits in IBRs can significantly affect system stability and introduce limit-induced bifurcations. Exploring more precise control and protection effects in IBR-dominated systems in combination with load recovery models is also a relevant research direction.

\bibliographystyle{IEEEtran}
\bstctlcite{IEEEexample:BSTcontrol}
\bibliography{references}

\begin{thebibliography}{10}
\providecommand{\url}[1]{#1}
\csname url@samestyle\endcsname
\providecommand{\newblock}{\relax}
\providecommand{\bibinfo}[2]{#2}
\providecommand{\BIBentrySTDinterwordspacing}{\spaceskip=0pt\relax}
\providecommand{\BIBentryALTinterwordstretchfactor}{4}
\providecommand{\BIBentryALTinterwordspacing}{\spaceskip=\fontdimen2\font plus
\BIBentryALTinterwordstretchfactor\fontdimen3\font minus
  \fontdimen4\font\relax}
\providecommand{\BIBforeignlanguage}[2]{{%
\expandafter\ifx\csname l@#1\endcsname\relax
\typeout{** WARNING: IEEEtran.bst: No hyphenation pattern has been}%
\typeout{** loaded for the language `#1'. Using the pattern for}%
\typeout{** the default language instead.}%
\else
\language=\csname l@#1\endcsname
\fi
#2}}
\providecommand{\BIBdecl}{\relax}
\BIBdecl

\bibitem{milano2018foundations}
F.~Milano, F.~D{\"o}rfler, G.~Hug, D.~J. Hill, and G.~Verbi{\v{c}},
  ``Foundations and challenges of low-inertia systems,'' in \emph{2018 Power
  Systems Computation Conference (PSCC)}.\hskip 1em plus 0.5em minus
  0.4em\relax IEEE, 2018, pp. 1--25.

\bibitem{markovic2021understanding}
U.~Markovic, O.~Stanojev, P.~Aristidou, E.~Vrettos, D.~Callaway, and G.~Hug,
  ``Understanding small-signal stability of low-inertia systems,'' \emph{IEEE
  Trans. on Power Systems}, vol.~36, no.~5, pp. 3997--4017, 2021.

\bibitem{gross2019effect}
D.~Gro{\ss}, M.~Colombino, J.-S. Brouillon, and F.~D{\"o}rfler, ``The effect of
  transmission-line dynamics on grid-forming dispatchable virtual oscillator
  control,'' \emph{IEEE Trans. on Control of Network Systems}, vol.~6, no.~3,
  pp. 1148--1160, 2019.

\bibitem{henriquez2020grid}
R.~Henriquez-Auba, J.~D. Lara, C.~Roberts, and D.~S. Callaway, ``Grid forming
  inverter small signal stability: Examining role of line and voltage
  dynamics,'' in \emph{IECON 2020 The 46th Annual Conference of the IEEE
  Industrial Electronics Society}.\hskip 1em plus 0.5em minus 0.4em\relax IEEE,
  2020, pp. 4063--4068.

\bibitem{concordia1982load}
C.~Concordia and S.~Ihara, ``Load representation in power system stability
  studies,'' \emph{IEEE Trans. on Power Apparatus and Systems}, no.~4, pp.
  969--977, 1982.

\bibitem{kosterev1999model}
D.~N. Kosterev, C.~W. Taylor, and W.~A. Mittelstadt, ``{Model validation for
  the August 10, 1996 WSCC system outage},'' \emph{IEEE Trans. on Power
  Systems}, vol.~14, no.~3, pp. 967--979, 1999.

\bibitem{milanovic2012international}
J.~Milanovic, K.~Yamashita, S.~Villanueva, S.~Djokic, and L.~Korunovi{\'c},
  ``International industry practice on power system load modeling,'' \emph{IEEE
  Trans. on Power Systems}, vol.~28, no.~3, pp. 3038--3046, 2012.

\bibitem{arif2017load}
A.~Arif, Z.~Wang, J.~Wang, B.~Mather, H.~Bashualdo, and D.~Zhao, ``Load
  modeling—a review,'' \emph{IEEE Transactions on Smart Grid}, vol.~9, no.~6,
  pp. 5986--5999, 2017.

\bibitem{emadi2006constant}
A.~Emadi, A.~Khaligh, C.~H. Rivetta, and G.~A. Williamson, ``Constant power
  loads and negative impedance instability in automotive systems: definition,
  modeling, stability, and control of power electronic converters and motor
  drives,'' \emph{IEEE Trans. on Vehicular Technology}, vol.~55, no.~4, pp.
  1112--1125, 2006.

\bibitem{allen2000interaction}
E.~H. Allen and M.~Ilic, ``Interaction of transmission network and load phasor
  dynamics in electric power systems,'' \emph{IEEE Trans. on Circuits and
  Systems I: Fundamental Theory and Applications}, vol.~47, no.~11, pp.
  1613--1620, 2000.

\bibitem{d2015virtual}
S.~D’Arco, J.~A. Suul, and O.~B. Fosso, ``A virtual synchronous machine
  implementation for distributed control of power converters in smartgrids,''
  \emph{Electric Power Systems Research}, pp. 180--197, 2015.

\bibitem{ajala2021model}
O.~Ajala, M.~Lu, S.~Dhople, B.~B. Johnson, and A.~Dominguez-Garcia, ``Model
  reduction for inverters with current limiting and dispatchable virtual
  oscillator control,'' \emph{IEEE Trans. on Energy Conversion}, 2021.

\bibitem{johnson2022generic}
B.~B. Johnson, T.~G. Roberts, O.~Ajala, A.~D. Dom{\'\i}nguez-Garc{\'\i}a, S.~V.
  Dhople, D.~Ramasubramanian, A.~Tuohy, D.~Divan, and B.~Kroposki, ``A generic
  primary-control model for grid-forming inverters: Towards interoperable
  operation \& control.'' in \emph{HICSS}, 2022, pp. 1--10.

\bibitem{milano2010power}
F.~Milano, \emph{Power system modelling and scripting}.\hskip 1em plus 0.5em
  minus 0.4em\relax Springer, 2010.

\bibitem{khalilbook}
H.~K. Khalil, \emph{{Nonlinear systems; 3rd ed.}}\hskip 1em plus 0.5em minus
  0.4em\relax Prentice-Hall, 2002.

\bibitem{marszalek2005singularity}
W.~Marszalek and Z.~W. Trzaska, ``Singularity-induced bifurcations in
  electrical power systems,'' \emph{IEEE Trans. on Power Systems}, vol.~20,
  no.~1, pp. 312--320, 2005.

\bibitem{chow1995systems}
J.~H. Chow, P.~V. Kokotovic, and R.~J. Thomas, \emph{Systems and control theory
  for power systems}.\hskip 1em plus 0.5em minus 0.4em\relax Springer Science
  \& Business Media, 1995.

\bibitem{krause2013analysis}
P.~C. Krause, O.~Wasynczuk, S.~D. Sudhoff, and S.~D. Pekarek, \emph{Analysis of
  electric machinery and drive systems}.\hskip 1em plus 0.5em minus 0.4em\relax
  John Wiley \& Sons, 2013.

\bibitem{mahmoudi2016new}
A.~Mahmoudi, S.~Hosseinian, M.~Kosari, and H.~Zarabadipour, ``A new linear
  model for active loads in islanded inverter-based microgrids,'' \emph{Int.
  Journal of Elec. Power \& Energy Systems}, vol.~81, pp. 104--113, 2016.

\end{thebibliography}

\end{document}